\documentclass[conference]{IEEEtran}
\IEEEoverridecommandlockouts
\usepackage{amsmath}

\usepackage[utf8]{inputenc} 
\usepackage[T1]{fontenc}    
\usepackage{hyperref}       
\hypersetup{draft}
\usepackage{url}            
\usepackage{booktabs}       
\usepackage{comment}
\usepackage{amsfonts}       
\usepackage{nicefrac}  
\usepackage{amsthm}
\usepackage{microtype}      
\usepackage{xcolor}         
\usepackage{amsmath}
\usepackage{bbm}
\usepackage{interval}
\usepackage{dsfont}

\usepackage{mathtools}
\usepackage{booktabs}
\usepackage{caption}
\usepackage{subcaption}
\usepackage[ruled,vlined]{algorithm2e}
\usepackage{graphicx}

\usepackage{amsfonts}
\usepackage{cite}
\DeclareMathOperator{\E}{\mathbb{E}}
\DeclareMathOperator{\R}{\mathbb{R}}

\graphicspath{ {./} }
\usepackage{caption}
\usepackage{subcaption}
\usepackage{graphicx}
\graphicspath{ {./} }
\newcommand{\overbar}[1]{\mkern 1.5mu\overline{\mkern-1.5mu#1\mkern-1.5mu}\mkern 1.5mu}
\usepackage{amssymb}

\usepackage{diagbox}

\def\BibTeX{{\rm B\kern-.05em{\sc i\kern-.025em b}\kern-.08em
    T\kern-.1667em\lower.7ex\hbox{E}\kern-.125emX}}

  \IEEEoverridecommandlockouts\IEEEpubid{\makebox[\columnwidth]{ 978-1-6654-3540-6/22~\copyright~2022 IEEE \hfill} \hspace{\columnsep}\makebox[\columnwidth]{ }}
\begin{document}

\title{UAV-Aided Multi-Community Federated Learning}

\makeatletter
\newcommand{\linebreakand}{%
  \end{@IEEEauthorhalign}
  \hfill\mbox{}\par
  \mbox{}\hfill\begin{@IEEEauthorhalign}
}
\makeatother
\author{
\IEEEauthorblockN{Mohamad Mestoukirdi $^{(1)(2) *}$, Omid Esrafilian$^{(1) *}$, David Gesbert$^{(1)}$, Qianrui Li$^{(2)}$}
\IEEEauthorblockA{$^{(1)}$ Communication Systems Department, EURECOM, Sophia Antipolis, France}
\IEEEauthorblockA{$^{(2)}$Mitsubishi Electric R\&D Centre Europe, Rennes, France}
\thanks{$^{*}$ Equal contribution.
\newline M. Mestoukirdi, Q. Li are with Mitsubishi Electric R\&D Centre Europe. 
Email: \{M.Mestoukirdi,Q.Li\}@fr.merce.mee.com
\newline M. Mestoukirdi, O. Esrafilian and D. Gesbert are with the Communication
Systems Department, EURECOM, Sophia-Antipolis, France. 
Emails: \{mestouki, esrafili, gesbert\}@eurecom.fr.
\newline Part of O. Esrafilian, and D. Gesbert work was funded via HUAWEI
France supported Chair on Future Wireless Networks at EURECOM.
}
}

\maketitle

\begin{abstract}
In this work, we investigate the problem of an online trajectory design for an Unmanned Aerial Vehicle (UAV) in a Federated Learning (FL) setting where several communities exist, each defined by a unique task to be learned. In this setting, spatially distributed devices belonging to each community collaboratively contribute towards training their community model via wireless links provided by the UAV. Accordingly, the UAV acts as a mobile orchestrator coordinating the transmissions and the learning schedule among the devices in each community, intending to accelerate the learning process of all tasks. We propose a heuristic metric as a proxy for the training performance of the different tasks. Capitalizing on this metric, a surrogate objective is defined which enables us to jointly optimize the UAV trajectory and the scheduling of the devices by employing convex optimization techniques and graph theory. The simulations illustrate the out-performance of our solution when compared to other handpicked static and mobile UAV deployment baselines.

\end{abstract}

\section{Introduction}

Thanks to their fast on-demand deployment and their inherent maneuvering capabilities, Unmanned Aerial Vehicles (UAVs) evolved to complement or even substitute static access points in multiple areas \cite{8011325}, ranging from emergency and disaster response scenarios \cite{malandrino2019planning} to flying relays that can expand the coverage area of ground base stations by acting as a relay to ground devices \cite{9162753}. More recently, the usage of UAVs to facilitate Federated Learning (FL) model training of ground and airborne units has gained significant attention. Federated Learning offers edge devices the opportunity to collaboratively train a model under the orchestration of a parameter server (PS) by iteratively aggregating locally optimized models without off-loading their local data\cite{mcmahan2017communication}. In \cite{9400376}, a UAV trajectory path planning problem has been formulated in order to govern the participation of the straggling devices during training. Their solution optimizes the UAV trajectory to balance the local model updates computation and transmission times at each learner to fit in each communication round time slot and guarantee the widest participation of devices. In \cite{DBLP:journals/corr/abs-2002-08196}, a joint power allocation and scheduling design is proposed to optimize the convergence rate of FL training among a swarm of UAVs. 

\noindent While orchestrating FL training is not a typical use-case for UAVs in urban areas --- where wireless connectivity is guaranteed and static Access points (APs) can act as robust front-haul orchestrators ---, it sounds appealing to deploy UAVs as FL orchestrators for Internet of things (IoT) devices in rural areas where a multitude of data are expected to be generated by massive numbers of machines and sensors. The generated data are envisioned to help in the management and optimization of the industrial and agricultural economy by serving as a training data feed for predictive machine learning models. In this setting, training models centrally by pooling the massive amounts of data from edge IoT devices may inflict a hit over their energy budget, especially if data are high dimensional. Additionally, the deployment of static APs in rural unpopulated areas is costly, and inefficient as they are mostly under-utilized. Moreover, the static nature of APs  does not guarantee a good wireless channel quality to the IoT edge devices, worsening their transmission times and consequently expanding their energy expenditure. Alternatively, devices can rely on decentralized machine learning (ML) schemes over wireless device-to-device (D2D) networks to train ML models \cite{DBLP:journals/corr/abs-2202-00955}. However, due to the limited communication range of edge devices, their connectivity is not always guaranteed. As a result, distributed D2D ML algorithms may perform poorly.

Unlike the previous works, in this paper, we investigate an online path planning problem of a UAV missioned to orchestrate the FL training among devices belonging to different communities. Each community consists of statistically heterogeneous  devices (i.e with non-IID datasets) that wish to train a model corresponding to their unique community task. The model updates are transmitted by the devices through a lossy channel, therefore, the successful participation of all devices during each training round is not guaranteed. Our goal is to establish learning fairness among the different community tasks$/$models during training, therefore, guaranteeing a desirable inference performance of all the different tasks at the end of training. This is achieved by employing a heuristically derived metric that is able to capture the training performance and the scheduling requirements of the different tasks throughout the course of training. Capitalizing on this metric, we devise a surrogate optimization problem which is solved by the UAV at each communication round, to dynamically schedule devices and optimize the UAV trajectory to successfully pool their model updates. Our solution aims at steering the scheduling and the UAV control in favor of users belonging to communities that are seen to lack behind in terms of convergence and as a result, establish learning fairness among all the tasks. To the best of our knowledge, this is the first work that addresses a Multi-Community FL setting, orchestrated by a UAV.

\section{System Model}
We consider a scenario where a UAV acts as flying orchestrator for FL training across different communities of devices in a service area. The considered area is composed of a total of $C$ communities. Each community $c$ consists of $|\mathcal{K}_c|$ ground devices where $\mathcal{K}_c$ is a set of  devices' index in community $c$ and $|.|$ denotes the cardinality function. The devices in each community wish to collaboratively train a supervised learning model to fit to their corresponding community task in a federated manner. We emphasize that the models that we wish to train at the different communities are unrelated, hence, there is no collaboration among devices which are not in the same community. We denote the total number of devices in the service area by $K$, such that $\sum_{c=1}^C | \mathcal{K}_c | = K $. We assume that each device $k$ is endowed with a training dataset $\mathcal{D}_k$, and a validation dataset relative to its corresponding community task. The $k$-th
ground device, is located at $ \bold{u}_k= \left[x_k, y_k\right]^T \in \R^2$. By no means, the ground level device assumption is restrictive and the proposed solution in this work can in principle be
applied to a scenario where the devices are located in 3D. 
The UAV's mission consists of $M$ communication rounds. During each communication round $m\in [1, M]$ the UAV collects the locally optimized models from the devices (yet to be optimally scheduled later) in different communities. At the end of each round the UAV aggregates the collected model updates from each community devices, to obtain new community-specific global models. Each global model is then broadcasted back to its' corresponding community devices, therefore initiating a new communication round.
The UAV is characterized by a battery budget which allows it to maneuver for a distance of $L_{total}$ meters with a constant velocity of $v \, \text{m/s}$. Moreover, the UAV is assumed to fly at an altitude $z(t)$ above the ground and the horizontal location of the UAV at time $t$ is denoted by ${\bold{v}(t)} = \left[x(t),y(t)\right]^T$. We assume that the UAV is equipped with a GPS, hence, its location is known at each time stamp $t$. For space limitation, we do not consider the optimization over the UAV altitude, and assume that the UAV flies at a fixed altitude $z(t) = H$. 
Since controlling the UAV in continuous time is cumbersome, we discretize each communication round into $N$ time steps. Hence, the UAV trajectory is defined by a set of discrete locations $\{\bold{v}[n]=\left[x[n],y[n]\right]^T, n\in [1, N] \}$, where each two consecutive UAV locations are connected with a straight line.
\subsection{Channel Model}
\label{CM}
\noindent We define the wireless channel gain between device $k$ and the UAV at time step $n$ as a log-normal fading channel under Additive White Gaussian Noise (AWGN) , given by : 
\begin{align}
h_{k,s}[n] = \frac{\beta_s}{ d_k[n]^{\alpha_s}}\xi_s, \label{eq:channel_model}
\end{align}
where $\xi_s$ denotes the shadowing component that is modeled as a Log-normal distribution $\xi_s \sim \text{Lognormal}(0, \sigma_s^2)$. $s\in\left\{ \text{LoS},\text{NLoS}\right\}$ emphasizes the strong dependence of the propagation parameters on the Line-of-Sight (LoS) or Non-Line-of-Sight (NLoS) segments.
$\beta_s$ is the average gain at the reference point $d = 1$ meter, and $d_k[n]= \sqrt{\|\bold{u_k} - \bold{v}[n] \|^2 + H^2} $ is the distance between the ground device $k$ and the UAV at step $n$.
\noindent The LoS event probability of the link between the UAV at time step $n$ and device $k$ is given by \cite{AlhKaLa} :
\begin{align} 
\rho_k[n] = \frac{1}{1+\exp(-a_1\theta_k[n] + a_2)},
\end{align}
where $\theta_k[n] = \arctan(\frac{H}{\|\bold{u_k} - \bold{v}[n] \|})$ is the elevation angle, parameters $\{a_1,a_2\}$ denote the model coefficients of the LoS probability which depends on the structure of the city and can be obtained according to \cite{AlhKaLa}.

Without loss of generality, we assume that the model updates are transmitted in packets across a lossy channel, and that the UAV has enough power to transmit the global models in the downlink (DL) for all devices with an average packet success rate equal to one, from every point inside the service area. In the next subsection, we derive the expression of the average Packet Error Rate (PER) experienced by the ground devices while transmitting their updates in the uplink (UL). 
\subsection{Average Packet Error Rate}
\noindent We define $q(\gamma)$, the instantaneous PER, representing the probability of packet detection error at a given signal-to-noise ratio (SNR) $\gamma$. We assume that the packets are erroneously detected with a probability $0\leq q(\gamma)\leq 1 $ if the instantaneous SNR resides below a threshold $\gamma_0$, and $q(\gamma) = 0$, otherwise.  The instantaneous SNR  experienced by the UAV when device $k$ is in $s\in\{\text{LoS},\,\text{NLoS}\}$ at step $n$, is given by :
\begin{align}
\gamma_{k,s}[n] = \frac{P_k\,h_{k,s}[n]}{\,N_0},
\end{align}
where $P_k$ is the transmission power of device $k$, and $N_0$ is the noise power level.
In accordance with \eqref{eq:channel_model}, $\gamma_{k,s}[n]$ follows a log-normally distribution $\gamma_{k,s}[n] \sim \text{Lognormal}(\mu_{k,s}[n], \sigma_s^2)$, where $\mu_{k,s}[n] = \log(\frac{P_k\,\beta_s}{N_0\,d_k[n]^{\,\alpha_s}})$. We denote by $f_{\gamma_{k,s}}(\gamma)$ the probability density function of $\gamma_{k,s}[n]$. 

The average PER experienced by device $k$ during the model transmission in the UL at UAV location at time step $n$ can be written as :
\begin{align}
\centering
&\bar{q}_{k}[n] = \E_{s}\left[\,\E_{\gamma_{k,s}[n]}\left[\,q(\gamma_{k,s}[n])\right]\,\right].
\label{average-per}
\end{align}
The inner expectation is over the instantaneous SNR randomness, while the outer expectation over $s$ is with respect to the channel LoS/NLoS segments probabilities. 
Hereafter we drop the time step index $n$ for ease of notation. For a given time step $n$,  averaging over the LoS/NLoS probabilities, we can rewrite (\ref{average-per}) as :
\begin{align} 
\bar{q}_k &\nonumber= \rho_k\,\E_{\gamma_{k,\tiny \text{LoS}}}\left[q(\gamma_{k,\tiny\tiny\text{LoS}})\right] + (1-\rho_k)\,\E_{\gamma_{k,\tiny\text{NLoS}}}\left[q(\gamma_{k,\tiny\tiny\text{NLoS}})\right] 
\\&\nonumber\overset{(a)}{\le }\rho_k\int_0^{\gamma_0}f_{\gamma_{k,\tiny \text{LoS}}}(\gamma)d\gamma + \overbar{\rho}_k \int_0^{\gamma_0}f_{\gamma_{k,\tiny \text{NLoS}}}(\gamma)d\gamma
\\&=\rho_k\,\,\phi\left(\gamma_0,\gamma_{k,\tiny\tiny\text{LoS}}\right) +\overbar{\rho}_k\,\,\phi\left(\gamma_0,\gamma_{k,\tiny\tiny\text{NLoS}}\right), \label{eq:per_upper_bound}
\end{align}
where $\overbar{\rho}_k = (1-\rho_k)$. Step $(a)$ holds given that $q(\gamma) \le 1,\, \forall \, \gamma \in \left(0,\gamma_0\right)$. 
$\phi\left(\gamma_0,\gamma_{k,s}\right) = \mathds{P}\left(\gamma_{k,s}<\gamma_0\right)$ is the cumulative density function of $\gamma_{k,s}$, and is written as :
\begin{align}
\phi\left(\gamma_0,\gamma_{k,s}\right) = \frac{1}{2}\left[1+\text{erf}\left(\,\frac{\log(\gamma_0)-\mu_{k,s}}{\sigma_{s}\sqrt{2}}\right)\right],
\end{align}
where $\text{erf}(x)$ is the error function. \section{Community FL and UAV trajectory Modelling}
In this section, we describe the Federated Learning training process across devices belonging to different communities.

\subsection{Classical Federated Learning}
\noindent In classical FL settings \cite{mcmahan2017communication}, the goal is to collaboratively train a model across different learners to find a
parameterized predictor $f_\vartheta:\mathcal{X}\to\mathcal{\hat{Y}}$ that minimizes the expected local risk over the learners datasets. Given a set of different communities and their corresponding tasks, this reflects as finding the predictor of each community $c$ that minimizes :
\begin{equation}
    F(\vartheta) = \sum_{k\in \mathcal{K}_c} p_k F_k(\vartheta)
    \label{aggregatedloss}
\end{equation}
where $F_k(\vartheta) = \E_{(x,y) \sim \mathcal{D}_k}\left[\mathcal{L}_c(f_\vartheta(x),y)\right]$ is the expected local risk of device $k \in \mathcal{K}_c$ with respect to its local dataset $\mathcal{D}_k$, and $\mathcal{L}_c :\mathcal{\hat{Y}}\times \mathcal{Y}\to \mathbb{R}^+$ denotes community $c$ loss function. $\vec{p}=(p_1,\dots,p_{|\mathcal{K}_c|})$ is a weighting scheme such that $\sum_{k\in\mathcal{K}_c}p_k = 1$.
In traditional FL training, a static orchestrator is missioned to supervise the training among the learners. Training occurs during multiple communication rounds, each composed of a DL, training and then an UL phase. During the DL phase, global models are transmitted to all learners, which in turn train them locally. Then, the learners transmit their model updates to the orchestrator in the UL phase. Subsequently, a new global model is produced and a new round is initiated.

\subsection{UAV-aided Orchestration}
\noindent In this work, and unlike traditional FL implementations, a mobile UAV is deployed to orchestrate the training of the different community tasks available. In this setting, a communication round starts as the UAV finds an optimized trajectory as well as schedules a set of devices from the different available communities to participate in the training. Then, the global models are broadcasted to all communities during the DL phase with a PER equal to zero, as explained in Sec. \ref{CM}. The models are then optimized locally at the scheduled devices. Model updates are then sent back in the UL phase, as the UAV maneuvers following the optimized path found earlier, while governing a favorable channel condition for the scheduled devices, and consequently, a low packet error transmission rate, to successfully gather their updates.
To limit the energy spent by the UAV during the UL phase, we limit the total distance that can be travelled by the UAV during each round to $L_{max}$ meters. We assume that the ground devices are served by the UAV in a Time-Division Multiple Access (TD-MA) manner in the UL, and that a maximum of $\overbar{K}$ devices can be served by the UAV at each time step. Note that each device can be scheduled at most once during each round.
\section{Accounting for the learning performance}
Accounting for the learning performance is essential to the online path planning optimization problem that we wish to solve. In a single community vanilla FL setting, the for-seen advantage of sampling a device during a communication round is proportional to its dataset sizes. However, this is partially true in a multi-community FL setting, as scheduling (i.e device sampling) and resource allocation should be carried out while accounting to the relative learning performance of the different available communities. Particularly in our case, scheduling and UAV trajectory planning should be considered to insure a low PER for devices with tasks that are seen to fall behind other communities in terms of convergence.

\noindent Theoretically, the convergence rate of the FL models can be quantified based on the level of heterogeneity of the datasets available at the learners, their participation rate, and the model architecture. Unfortunately, computing the convergence rates in practice is not trivial, especially in settings where the loss landscape is non convex and datasets are heterogeneously distributed\cite{DBLP:journals/corr/abs-1812-06127}. Consequently, we choose the Coefficient of Variation (CoV), computed periodically during training, over the average validation accuracy of each community devices, as a metric of choice, to quantify the training performance of each community model. 

\noindent The motivation behind using the CoV is its ability to capture the current model performance difference among the devices belonging to the same community, compared to their average performance, ergo convey how well the current model performs over the devices' local datasets. Consequently, the CoV being calculated periodically offers a measure of goodness of the available different community models during training, which we can rely on to compare the training performance of the underlying tasks and quantify their scheduling and resource allocation requirements. 



\noindent The CoV of each community $c\,$, is updated by the UAV every $\ell$ communication rounds, and is given by
\begin{equation}
\psi_c  = \frac{\sqrt{\sum_{k \in \mathcal{ K}_c}\left(\varepsilon_{k} - \overbar{\varepsilon}_{\hspace{0.05cm}c} \right)^2 }}{\overbar{\varepsilon}_{\hspace{0.05cm}c}}\quad \forall c,
\end{equation}
\noindent where $\varepsilon_{k}$ is the average validation accuracy for device $k$ at each community which is computed over $\ell$ rounds as follows :
\begin{equation}
\varepsilon_{k} = \frac{1}{\ell} \sum_{j=m - \ell}^{m-1}\varepsilon_k(\vartheta_j).
\end{equation}
\noindent $\varepsilon_k(\vartheta_j)$ is the validation accuracy computed locally at device $k$ over the validation sample set, using the global model parameterized by $\vartheta_j$ transmitted in the DL during round $j$, and $\overbar{\varepsilon}_{\hspace{0.05cm}c}$ is the weighted average validation accuracy over all devices in community $c$ which is given by :
\begin{equation} 
\overbar{\varepsilon}_{\hspace{0.05cm}c} = \sum_{k\in \mathcal{K}_c} p_k\, \varepsilon_k,\text{ such that} \,\,\, p_k = \frac{|\mathcal{D}_k|}{\sum_{i\in \mathcal{K}_c}|\mathcal{D}_i|},
\label{eq.10}
\end{equation}
where $|\mathcal{D}_k|$ is the training data set size of device $k$. 

\noindent We assume that $\varepsilon_k$ is transmitted alongside the local models during the UL phase at each round by the scheduled devices. However, if $\varepsilon_k$ is not received during the round in which the CoV is updated, the last successfully received value is considered to update the CoV.


\section{UAV Trajectory Planning}
In this section, we seek to find an optimized UAV trajectory during each communication round, in order to improve the overall learning performance within the communities.

\noindent We introduce a surrogate optimization problem which enables us to optimize the UAV trajectory for collecting the model updates from a subset of devices of each community to improve the performance of learning. We define the surrogate optimization problem at each communication round as follows
\begin{subequations}\label{eq:surrogate_opt_problem}
\begin{align}
  \begin{split}
           \max_{\substack{\mathcal{V}, \mathcal{W}}
} & \quad \sum_{n\in[1, N]} 
\sum_{k\in[1, K]} \omega_{k}[n] \, (1-\bar{q}_k[n]) \, \delta_k
 \end{split}\\
 \begin{split}
          \text{s.t.}&\quad \sum_{n\in[1, N]} \omega_{k}[n] \leq 1 , \forall k, \label{eq:P1_one_per_round}
 \end{split}\\
 \begin{split}
          &\quad \sum_{k\in[1, K]} \omega_{k}[n] \leq \overbar{K} , \forall n, \label{eq:P1_max_connected_ue}
 \end{split}\\
 \begin{split}
          &\quad \sum_{n=1}^{N-1} \| {{\bf{v}}}[n+1] - {{\bf{v}}}[n]\|\leq L_{max}, \label{eq:P1_max_len} 
 \end{split}\\ 
 \begin{split}
          &\quad  {{\bf{v}}}[1] = {{\bf{v}}}_{\text{I}}, \label{eq:P1_start_point} 
 \end{split}
\end{align}
\end{subequations}
\noindent where $\mathcal{V} = \{{{\bf{v}}}[n], \forall n \}$ is the UAV trajectory, $\mathcal{W} = \{ \omega_{k}[n] \in \{0, 1\}, \forall n, k\}$ is a set of scheduling binary variables where $\omega_{k}[n]$ indicates that device $k$ is scheduled at time step $n$ if one, and otherwise if it is zero.  Constraint \eqref{eq:P1_one_per_round} implies that a device can only be served once by the UAV at each communication round, and \eqref{eq:P1_max_connected_ue} indicates the maximum of $\overbar{K}$ devices can be served by the UAV at each time step. Constraint \eqref{eq:P1_max_len} is the maximum length of the UAV trajectory allowed in each round, and ${{\bf{v}}}_{\text{I}}$ is the staring location at each round (i.e. ${{\bf{v}}}_{\text{I}}$ can be the location of the UAV at the end of the previous communication round to guarantee a continuous trajectory throughout the entire mission). $\delta_k$ captures the importance of  participation of device $k$ during the current round. $\delta_k$ is a function of device $k$ weight $p_k$ given in (\ref{eq.10}), and the CoV of the community which it belongs to $\psi_c$. Moreover, in order to guarantee fairness over the participation of devices throughout the course of training, we impose an extra weight $\lambda$ $\left(\lambda > 1\right)$, for devices that have failed to transmit their model updates successfully, or have not been scheduled during the previous round. Hence, the importance of scheduling device $k \in \mathcal{K}_c$ at each round is given by
\begin{equation}
    \delta_k =
    \begin{cases}
    p_k\,\psi_c\,\lambda , & \text{\small if $\forall\, n, $ $\,w_k[n] = 0$ during the previous round,} \\
    p_k\,\psi_c, & \text{\small Otherwise.} 
    \end{cases}
    \label{imp}
\end{equation}
Solving problem \eqref{eq:surrogate_opt_problem} is challenging since the exact close form of $\bar{q}_k[n]$ is not available. To solve this problem we first simplify the objective function by finding an approximate for $\bar{q}_k[n]$. Since  $\bar{q}_k[n]$ comprises the $\text{erf}(.)$ function, a close approximation can be obtained by using the logistic function. Therefore, an approximate for $\bar{q}_k[n]$ is given by
\begin{equation}
\bar{q}_k[n] \approx \Tilde{q_k}[n] \triangleq \frac{1}{1+\exp(b_1\theta_k[n] + b_2)},
\end{equation}
\noindent where $\theta_k[n]$ is the elevation angle between the UAV at time step $n$ and the $k$-th device. The parameters $\{b_1, b_2\}$ can be found using regression techniques on the samples taken from \eqref{eq:per_upper_bound} for different UAV and device locations. For further simplification, we also relax the binary scheduling variables $\mathcal{W}$ into continuous variables. Hence, problem \eqref{eq:surrogate_opt_problem} by substituting $\Tilde{q_k}[n]$ and relaxed scheduling variables can be reformulated as follows
\begin{subequations}\label{eq:approximate_opt_problem}
\begin{align}\small
  \begin{split}
           \max_{\substack{\mathcal{V}, \mathcal{W}}
} &  \sum_{n\in[1, N]} 
\sum_{k\in[1, K]} \omega_{k}[n] \, (1-\frac{1}{1+\exp(b_1\theta_k[n] + b_2)}) \, \delta_k
 \end{split}\\
 \begin{split}
          \text{s.t.}&\quad \eqref{eq:P1_one_per_round}, \eqref{eq:P1_max_connected_ue}, \eqref{eq:P1_max_len}, \eqref{eq:P1_start_point},
 \end{split}\\
  \begin{split}
          &\quad 0 \leq \omega_k[n] \leq 1, \forall n, k. \label{eq:P2_relaxed_omega}
 \end{split}
\end{align}
\end{subequations}
\noindent However, having simplified the objective function, this problem is still difficult to solve as it is a non-convex/concave mixed-integer optimization problem. To tackle this difficulty, we split the optimization problem  \eqref{eq:approximate_opt_problem} into two sub-problems of device scheduling and UAV trajectory optimization. In the first phase, the devices are scheduled while fixing the UAV trajectory. Then in the second phase,  given the scheduled devices from the first phase the UAV trajectory is optimized. The algorithm iterates between two phases until convergence.
\subsection{Device Scheduling}
\label{devicescheduling}
\noindent For a given UAV trajectory $\mathcal{V}$, the ground device scheduling can be optimized as follows
\begin{subequations}
\label{eq:user_scheduling}
\begin{align}
  \begin{split}
           \max_{\substack{\mathcal{W}}
} & \quad \sum_{n\in[1, N]} 
\sum_{k\in[1, K]} \omega_{k}[n] \, (1-\Tilde{q_k}[n]) \, \delta_k
 \end{split}\\
 \begin{split}
          \text{s.t.}&\quad \eqref{eq:P1_one_per_round}, \eqref{eq:P1_max_connected_ue}, \eqref{eq:P2_relaxed_omega}.
 \end{split}
\end{align}
\end{subequations}
\noindent This problem is a standard Linear Program (LP) and can be
solved by using any optimization tools such as CVX \cite{CVX}.
\subsection{Trajectory Optimization}
\noindent Having optimized the scheduling variables $\mathcal{W}$, the optimal UAV trajectory can be obtained by solving the following optimization
\begin{subequations}\label{eq:trajectory_opt_problem}
\begin{align}\small
  \begin{split}
           \max_{\substack{\mathcal{V}}
} &  \sum_{n\in[1, N]} 
\sum_{k\in[1, K]} \omega_{k}[n] \, (1-\frac{1}{1+\exp(b_1\theta_k[n] + b_2)})\, \delta_k
 \end{split}\\
 \begin{split}
          \text{s.t.}&\quad \eqref{eq:P1_max_len},  \eqref{eq:P1_start_point}.
 \end{split}
\end{align}
\end{subequations}
This problem is still non-convex. By introducing slack variables $\mathcal{S}=\{ S_k[n], \forall n, k \}$, $\mathcal{T}=\{ \theta_k[n], \forall n, k \}$, and $\mathcal{R}=\{ r_k[n], \forall n, k \}$ problem \eqref{eq:trajectory_opt_problem} can be rewritten as  
\begin{subequations}\label{eq:trajectory_equal_opt_problem}
\begin{align}
  \begin{split}
           \max_{\substack{\mathcal{V}, \mathcal{S}, \mathcal{T}, \mathcal{R}}
} &  \sum_{n\in[1, N]} 
\sum_{k\in[1, K]} \omega_{k}[n] \, (1-\frac{1}{1+S_k[n]}) \, \delta_k \label{eq:P3_objective}
 \end{split}\\
  \begin{split}
          \text{s.t.}&\quad S_k[n] \leq \exp(b_1\theta_k[n] + b_2), \forall n, k,\label{eq:P3_S}
 \end{split}\\
  \begin{split}
          &\quad \theta_k[n] \leq \arctan(\frac{H}{r_k[n]}), \forall n, k,\label{eq:P3_theta}
 \end{split}\\
  \begin{split}
          &\quad r_k[n] = \| {{\bf{v}}}[n] - {{\bf{u}}}_k \|, \forall n, k, \label{eq:P3_R}
 \end{split}\\
 \begin{split}
          &\quad \eqref{eq:P1_max_len}, \eqref{eq:P1_start_point}.
 \end{split}
\end{align}
\end{subequations}
\noindent Without loss of optimality the constraints \eqref{eq:P3_S} and \eqref{eq:P3_theta} can be met with equality. It can be verified that objective function \eqref{eq:P3_objective} is a concave function for $S_k[n] \geq 0 $, however, problem \eqref{eq:trajectory_equal_opt_problem} is still non-convex/concave. To solve this problem efficiently, we employ the sequential convex programming techniques by using a local first-order Taylor estimation to convert the problem into a convex form. To do so, it can be shown that the right hand side functions in constraints \eqref{eq:P3_S}, \eqref{eq:P3_theta}, and \eqref{eq:P3_R} are convex functions of $\theta_k[n]$, $r_k[n]$, and ${{\bf{v}}}[n]$, respectively, when $\theta_k[n], r_k[n] \geq 0$. Since every convex function can be lower-bounded by its first-order Taylor approximation, a lower bound of problem \eqref{eq:trajectory_equal_opt_problem} is given by

 \begin{subequations}\label{eq:trajectory_opt_problem_lower_bound}
\begin{align}
  \begin{split}
           \max_{\substack{\mathcal{V}, \mathcal{S}, \mathcal{T}, \mathcal{R}}
} &  \sum_{n\in[1, N]} 
\sum_{k\in[1, K]} \omega_{k}[n] \, (1-\frac{1}{1+S_n[k]}) \, \delta_k \label{eq:P3_objective}
 \end{split}\\
  \begin{split}
          \text{s.t.}&\quad S_k[n] \le \Tilde{S}(\theta_k[n]), \forall n, k,
 \end{split}\\
  \begin{split}
          &\quad \theta_k[n] \le \Tilde{\theta}(r_k[n]), \forall n, k,
 \end{split}\\
  \begin{split}
          &\quad r_k[n] \geq \Tilde{r}({{\bf{v}}}[n]), \forall n, k,
 \end{split}\\
  \begin{split}
          &\quad S_n[k], \theta_k[n], r_k[n] \geq 0 , \forall n, k, 
 \end{split}\\
 \begin{split}
          &\quad \eqref{eq:P1_max_len},  \eqref{eq:P1_start_point}.
 \end{split}
\end{align}
\end{subequations}
\noindent where $\Tilde{S}(\theta_k[n]), \Tilde{\theta}(r_k[n])$, and $\Tilde{r}({{\bf{v}}}[n])$ are the local first-order Taylor approximation of functions in the right hand side of constraints \eqref{eq:P3_S}, \eqref{eq:P3_theta}, and \eqref{eq:P3_R} with respect to  $\theta_k[n], r_k[n]$, and ${{\bf{v}}}[n]$, respectively.

\subsection{Overall Algorithm and Convergence}
\noindent According to the preceding analysis, now we propose an iterative algorithm to solve the optimization problem \eqref{eq:approximate_opt_problem} by applying the alternating optimization method. We split the problem into two phases i) user scheduling, and ii) UAV trajectory optimization. In the first phase, the devices are scheduled while keeping the UAV trajectory fixed. Then in the second phase, given the optimized scheduling variables $\mathcal{W}$ from the first phase, the UAV trajectory is optimized. The algorithm iterates between two phases until convergence. Moreover, the obtained solution in each iteration is used as the input for the next iteration. The convergence of the aforementioned algorithm is guaranteed since, at each phase of device scheduling and the UAV trajectory optimization, the objective function is optimized and does not decrease compared to the previous phase which results in convergence to at least a local optima. The details of the proof are omitted for the sake of the limited space.

\subsection{Trajectory Initialization}
\noindent Due to the non-convexity/concavity of problem \eqref{eq:approximate_opt_problem}, the iterative solution proposed above will converge to a local minima. Therefore, it is of a crucial importance to suitability initialize the UAV trajectory. To do so, we use a low complexity graph-based algorithm to find a good candidate for the initial UAV trajectory. We define the graph $G(\mathcal{N}, \mathcal{E})$ comprising a set of nodes $\mathcal{N}$ and a set of edges $\mathcal{E}$. The graph nodes includes a set of  UAV locations where the UAV can fly to and is defined as $\mathcal{N} = \left\{ {\boldsymbol{\nu}}_k = [x_k, y_k, H]^T, k\in [1, K] \right \}$. This implies that the UAV has to fly to the location at top of the devices at a fixed altitude $H$. We also add ${{\bf{v}}}_{\text{I}}$ as node zero to $\mathcal{N}$. The graph edges consists of all the possible combinations of the segments between the nodes and is defined as $\mathcal{E} = \left\{ e_{i,j} = ({\boldsymbol{\nu}}_i, {\boldsymbol{\nu}}_j), {\boldsymbol{\nu}}_i, {\boldsymbol{\nu}}_j\in \mathcal{N}, {\boldsymbol{\nu}}_i \neq {\boldsymbol{\nu}}_j \right \}$. We assign a reward $r_{i, j}$ to each edge $e_{i,j}$ of the graph defined
\begin{align*}\small
 \begin{split}
    r_{i, j} := \max_{\substack{\omega_k[j], k\in [1, K]}
} & \quad 
\sum_{k\in[1, K]} \omega_{k}[j] \, (1-\Tilde{q}_k[j]) \, \delta_k
 \end{split}\\
 \begin{split}
          \text{s.t.}&\quad \sum_{k\in[1, K]} \omega_{k}[j] \leq \overbar{K},
 \end{split}
\end{align*}
where the index $j$ indicates when the UAV is at location ${\boldsymbol{\nu}}_j \in \mathcal{N}$. Then a trajectory is defined as set of connected edges starting from ${{\bf{v}}}_{\text{I}}$ in graph $G$ that maximized the sum rewards while satisfying the maximum UAV trajectory length constraint \eqref{eq:P1_max_len} and the constraint \eqref{eq:P1_one_per_round}. To solve this problem, a greedy algorithm is used where an optimized initial trajectory is found within the graph iteratively.
\section{Experiments}
In our simulations, the UAV flies at constant altitude $H=60\,m$ with a constant velocity $v = 20\,m/s$, a travel budget $L_{total} = 40\,km$, and $L_{max} = 800\,m$. The true propagation parameters are chosen similar to \cite{9162753}. The transmission power for all ground devices is set to $-20$ dB, with a noise level of $-95$ dB. We chose $\overbar{K} = 3$, $\gamma_0 = 10$, the periodicity of CoV updates $\ell = 4$, $\lambda= 1.5$. We  set $\psi_c = 1 \,\, \forall\, c$ at the first round.

We consider a sub-urban area of size $800 \times 800 \,m^2$, containing two communities $(C=2)$, of 6 devices each, defined by two different tasks to train, namely the CIFAR10 \cite{krizhevsky2009learning} and MNIST \cite{deng2012mnist} image classification tasks. We distribute the devices randomly inside the service area, and as a data partitioning strategy, and to enforce heterogeneity among the datasets, we randomly assign $2$
different label IDs to each member across the different communities as in \cite{DBLP:journals/corr/McMahanMRA16}. Then, we randomly and equally divide the samples corresponding to each label across devices
which own that label.
For both tasks. We use Fed-Prox \cite{li2020federated} with parameter $\mu = 0.1$, to tackle the heterogeneity burden induced by the partial participation and the data heterogeneity of the devices. The SGD optimizer is used with a fixed learning rate of 0.01, and momentum = 0.9. The batch size is set to 16, and number of epochs is set to 1. 

\noindent In addition to our proposed solution, we analyze 4 different, handpicked static and mobile deployments :
\begin{itemize}
    \item \textbf{\small A static UAV hovering at the Barycenter} of the devices emulating a BS deployment. Given that the UAV hovers still, we assume that each round lasts for 5 seconds in this particular experiment, which accounts to 100 meters traveled distance per round. 
    \item \textbf{\small A rectangular UAV trajectory (Fig. \ref{fig:rect}):} where the UAV attempts to cover the whole service area during its mission. Communication rounds are initiated over a set of hovering locations scattered on the predefined rectangular trajectory.
    \item \textbf{\small Optimal UAV control with Naive scheduling (No-CoV):} where the UAV attempts to maximize the objective in \eqref{eq:surrogate_opt_problem} while setting $\delta_k= p_k$ if device $k$ participated in the previous round, and $\delta_k =p_k\,\lambda$ otherwise.
    \item \textbf{\small Ideal case:} Representing the maximum achievable performance in an ideal FL setting.
\end{itemize}


\noindent In all experiments, the devices are scheduled akin to Sec. \ref{devicescheduling}. Moreover, in all deployments (excluding the Barycenter case), we consider that the duration of the DL/UL transmissions and local training is negligible compared to the maneuvering time taken by the UAV at each round.

\noindent In Fig. \ref{fig::accuracies}, we report the average validation accuracy attained by both tasks over Monte-Carlo (MC) simulations. At each MC iteration, the devices are distributed randomly. As expected, our solution achieves the highest average validation accuracy, compared to the other benchmarks. In the Rectangular trajectory case, the UAV energy budget is wasted on traversing predefined paths which does not take into account the exact devices' locations. Accordingly, the UL packet transmissions endure high average PER, resulting in low participation count during each round, undermining the convergence rate and hampering the training performance. In the Barycenter case, despite that the UAV hovers still at the mean devices location, the channel yet imposes a strong PER penalty over the devices packets transmission, especially for devices that reside far from the UAV, given their low transmission power. The poor performance of those two benchmarks is mainly related to the wrong UAV placement. Hence, in order to quantify the gain of our scheduling algorithm incorporating the CoV of the different communities, we devise the No-CoV experiment, in which the UAV attempts to maximize the objective in (\ref{eq:approximate_opt_problem}) while naively assigning the importance of the devices as a function of their dataset sizes, and ignoring their tasks training performance that is quantified by the CoV. As expected, employing the CoV in our optimization leads to faster convergence and a percentage gain of 14\% for the CIFAR10 task compared to the No-CoV experiment, while maintaining a similar performance in the MNIST case. This advantage stems from the inherent ability of the CoV in quantifying the training performance of the two different communities throughout the course of training, ergo enabling the UAV to establish learning fairness among them by prioritizing the CIFAR10 task in terms of device scheduling and trajectory planning, which is well recognized as a more complex task compared to the MNIST task.

\begin{figure}[t]
\centering
    \begin{subfigure}[t]{0.24\textwidth}
         \centering
         \includegraphics[width=\textwidth]{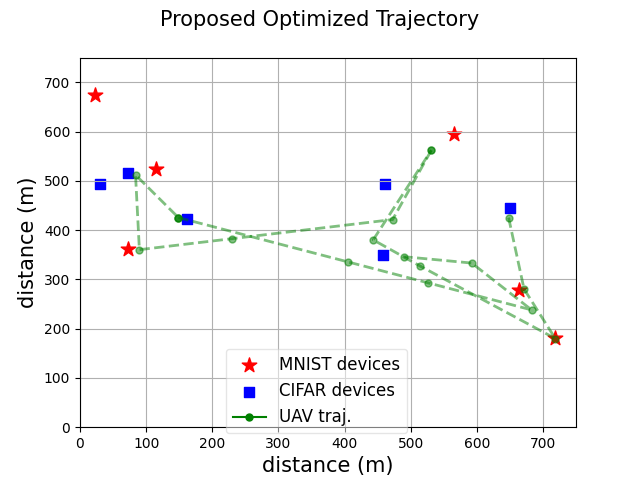}
         \centering
         \caption{\small      \centering Snapshot: Optimized UAV Trajectory (3 comm. rounds)}
         \label{fig:optimized}
     \end{subfigure}
      \begin{subfigure}[t]{0.24\textwidth}
         \centering
         \includegraphics[width=\textwidth]{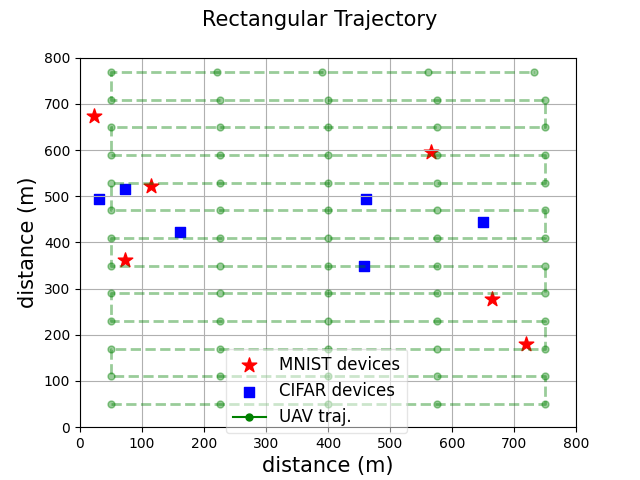}
         \caption{\small\centering Snapshot: Rectangular Trajectory}
         \label{fig:rect}
     \end{subfigure}
     \caption{\small\centering Optimized UAV Trajectory  vs Rectangular trajectory}
\end{figure}
\begin{figure}[t]
\centering
    \begin{subfigure}[t]{0.24\textwidth}
         \centering
         \includegraphics[width=\textwidth]{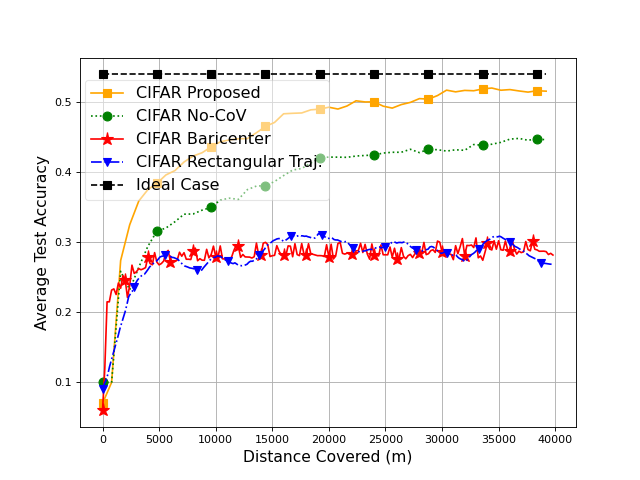}
         \centering
         \caption{\small CIFAR10 Test Accuracy}
         \label{fig:EMNIST_labelsol}
     \end{subfigure}
      \begin{subfigure}[t]{0.24\textwidth}
         \centering
         \includegraphics[width=\textwidth]{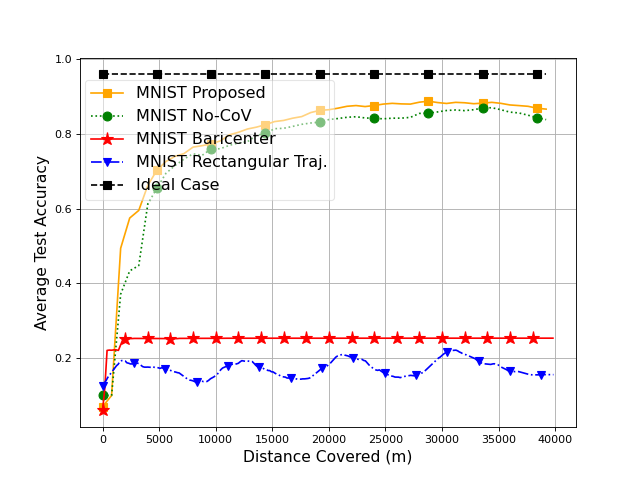}
         \caption{\small MNIST Test Accuracy}
         \label{fig:EMNIST_covsol}
     \end{subfigure}
     \caption{\small\centering Average Validation Accuracy attained by different strategies}
     \label{fig::accuracies}
\end{figure}
\section{Conclusion}
We studied the problem of an online path planning for a UAV missioned to orchestrate the training of different communities' tasks. We proposed a heuristic metric that is able to quantify the training performance and the scheduling requirements of the different tasks. Hinging on this metric, we devise a surrogate optimization problem which we solve iteratively using Convex optimization techniques, to schedule devices and find the optimal trajectory to successfully pool their updates, while aiming at achieving learning fairness among the available tasks. The performance of the proposed algorithm was evaluated via simulations, which highlighted its advantage compared to other benchmarks. 

\bibliographystyle{IEEEtran}
\bibliography{conference_101719}


        


\end{document}